\documentclass{article}
\usepackage[english]{babel}
\usepackage{float}
\usepackage{booktabs}
\usepackage{graphicx}
\usepackage{dcolumn}
\usepackage{bm}
\usepackage{cite}
\usepackage{amsthm} 
\include{cip-v3-pack-and-defined}

\usepackage[letterpaper,top=2cm,bottom=2cm,left=3cm,right=3cm,marginparwidth=1.75cm]{geometry}

\usepackage{amsmath}
\usepackage{graphicx}
\usepackage[colorlinks=true, allcolors=blue]{hyperref}
\usepackage{authblk}

\title{Inconsistency among evaluation metrics in link prediction}
\author[a,1]{Yilin Bi}
\author[a,1]{Xinshan Jiao}
\author[b]{Yan-Li Lee}
\author[a,$\ast$]{Tao Zhou}
\affil[a]{CompleX Lab, School of Computer Science and Engineering, University of Electronic Science and Technology of China, Chengdu 610054, China}
\affil[b]{School of Computer and Software Engineering, Xihua University, Chengdu 610039, China}
\affil[1]{Yilin Bi and Xinshan Jiao contributed equally to this work.}
\affil[$\ast$]{To whom correspondence should be addressed: \href{zhutou@ustc.edu}{zhutou@ustc.edu}}

\begin{document}
\maketitle

\begin{abstract}
Link prediction is a paradigmatic and challenging problem in network science, which aims to predict missing links, future links and temporal links based on known topology. Along with the increasing number of link prediction algorithms, a critical yet previously ignored risk is that the evaluation metrics for algorithm performance are usually chosen at will. This paper implements extensive experiments on hundreds of real networks and 25 well-known algorithms, revealing significant inconsistency among evaluation metrics, namely different metrics probably produce remarkably different rankings of algorithms. Therefore, we conclude that any single metric cannot comprehensively or credibly evaluate algorithm performance. Further analysis suggests the usage of at least two metrics: one is the area under the receiver operating characteristic curve (AUC), and the other is one of the following three candidates, say the area under the precision-recall curve (AUPR), the area under the precision curve (AUC-Precision), and the normalized discounted cumulative gain (NDCG). In addition, as we have proved the essential equivalence of threshold-dependent metrics, if in a link prediction task, some specific thresholds are meaningful, we can consider any one threshold-dependent metric with those thresholds. This work completes a missing part in the landscape of link prediction, and provides a starting point toward a well-accepted criterion or standard to select proper evaluation metrics for link prediction.
\end{abstract}

\textbf{Keywords}: link prediction, evaluation metrics, inconsistency

\section{Introduction}
Network is a powerful tool to represent many complex social, biological and technological systems, and network science is an increasingly active interdisciplinary research domain \cite{barabasi2016, newman2018a}. Link prediction is one of the most productive branches of network science, which aims at estimating likelihoods of missing links, future links and temporal links, based on known topology \cite{zhou2011, wang2015, martinez2016, kumar2020, dicakaran2020, zhou2021, yyliu}. As many observed networks are incomplete or dynamically changing, link prediction can find direct applications in inferring missing or upcoming links, such as the inference of missing biological interactions \cite{csermely2013, ding2014,BW2022}, the online recommendation of friends and products \cite{aiello2012, linyuan2012}, and the prediction of future scientific discoveries \cite{nagarajan2015, krenn2023}. Link prediction can also be considered as a touchstone to evaluate network models \cite{zhou2012, zhou2015a, catala2018, chasemian2019}, because a better understanding of network formation will in principle lead to a more accurate prediction algorithm. In addition, link prediction can be an important step in solving some challenging problems, like network reconstruction \cite{peixoto2018, squartini2018} and sparse training \cite{zhangy2022}, or an essential reason for some impressive phenomena, like polarization \cite{santos2021} and information cocoons \cite{hou2023} in online social networks.   
	
A huge number of link prediction algorithms have been proposed recently (see some selected representatives \cite{hasan2006, kleinberg2007, newman2008, zhou2009, guimera2009, liu2010, menon2011, carclo2013, zhou2015b, panlm2016, pech2017, zhangm2018, benson2018, kovacs2019, kitsak2020, ghasemian2020, wangh2023}), and an accompanying question is how to evaluate algorithm performance. A standard procedure is to divide the observed links into a training set and a probe set, and to train model parameters by using only the information contained in the training set. An algorithm’s performance is then measured by the closeness between ground truth and the algorithm’s prediction. Many evaluation metrics have already been utilized to quantify the above-mentioned closeness, including the two very popular ones, namely AUC \cite{hanely1982, bradley1997} and Balanced Precision (BP) \cite{zhou2023}, the one with increasing popularity, say AUPR \cite{davis2006}, as well as some occasionally used ones, such as Precision \cite{herlocker2004}, Recall \cite{herlocker2004}, F1-measure \cite{herlocker2004}, Matthews Correlation Coefficient (MCC) \cite{mattews1975}, NDCG \cite{jarvelin2002, wangy2013}, AUC-Precision \cite{carclo2013}, and so on.  
	
Everyone should be immediately aware of the crucial role of evaluation metrics, however, discussions about how to choose metrics in link prediction are rare. Recently, a few scientists have conducted criticism on popular metrics. Yang, Lichtenwalter, and Chawla \cite{yangy2015} argued that, when evaluating link prediction performance, the precision-recall curve might provide better accuracy than the ROC curve. Saito and Rehmsmeier \cite{saito2015} pointed out that AUC is inadequate to evaluate the performance of imbalanced classification problem, while link prediction is a typical imbalanced classification problem as most real-world networks are sparse \cite{genio2011}. Menand and Seshadhri claimed that neither AUC nor AUPR can well characterize the algorithm performance in link prediction for sparse networks, and proposed a vertex-centric measure \cite{menand2024}. To overcome the shortcoming of AUC for imbalanced classification, Muscoloni, and Cannistraci \cite{carclo2021} designed a novel metric named the area under the magnified ROC curve (AUC-mROC), which assign remarkably high weights to top-ranked links. Zhou \textit{et al.} \cite{zhou2023, jiao2024} proposed a method to quantitatively measure the discriminating ability of any metric and showed that AUC, AUPR and NDCG have significantly higher discriminating abilities than other well-known metrics.

In despite of those studies on evaluation metrics, thus far, there is no criterion or standard to select evaluation metrics: some scientists are conditioned to follow popular metrics, while some others have their own niche preferences (see, for example, Table 1 of Ref. \cite{zhou2023}). Such fact reminds us of a even more fundamental question, that is, whether those evaluation metrics provide statistically consistent rankings of algorithms. If the answer is YES, then we can breathe easy since it is not a big deal in choosing metrics, while if the answer is NO, we have to reexamine the related literature because an algorithm being superior according to some metrics may be at a disadvantage for other metrics, and even worse, researchers who are too eager to get their works published may only report beneficial results from some metrics but ignore negative results from other metrics. This is not a groundless worry, as a recent large-scale experimental study has shown that a winner for one metric may be a loser for another metric \cite{carclo2023}.  
	
Unfortunately, we have not found any answers to the above question in the literature. In this paper, we intend to provide a direct answer by analyzing correlations between evaluation metrics based on 25 algorithms and hundreds of real-world networks. The answer is a clear NO, and further analysis arrives to four practical suggestions in the selection of metrics, which are presented in the last section of this paper. We believe those suggestions can be considered as a useful guide in choosing evaluation metrics, before a commonly recognized standard for metric selection that may appear in the future.


\section{Results}
Consider a simple network $G(V,E)$, where $V$ is the set of nodes, $E$ is the set of links, the directionalities and weights of links are ignored, and the multiple links or self links are not allowed. We assume that there are some missing links in the set of unobserved links $U-E$, where $U$ is the universal set containning all $|V|(|V|-1)/2$ potential links. The task of link prediction is to find out those missing links. However, as we do not known whether a link in $U-E$ is a missing link or a nonexistent link, to evaluate the algorithm's accuracy, a standard procedure is to use part of the links in $E$ to predict the other part. Practically, we randomly divide the set $E$ into a training set $E^{T}$ and a probe set $E^{P}$, use only information in $E^T$ to predict missing links, and approximately treat $E^P$ as positive samples (i.e., missing links) while $U-E$ as negative samples (i.e., nonexistent links). Intuitively speaking, an algorithm assigning higher likelihoods to positive samples and lower likelihoods to negative samples is considered to be a well-performed algorithm.

Consider the evaluation metrics $M_1$ and $M_2$, as well as a series of algorithms $A_1, A_2, \cdots, A_P$ (see figure 1 for an illustration for $P=5$), for an arbitrary network $G$, $M_1$ and $M_2$ will respectively give each algorithm an evaluation score. According to those scores, we can obtain two rankings of algorithms by $M_1$ and $M_2$ (see figure 1A), and then get the correlation of $M_1$ and $M_2$ by measuring the correlation of the two rankings. Using the same procedure, we can consider a number of networks $G_1, G_2, \cdots, G_Q$ to calculate the average correlation between any two evaluation metrics (see figure 1B and figure 1C for an illustration for $Q=3$). To reduce the possible fluctuations, we utilize up to $P=25$ algorithms and up to $Q=340$ real networks (see Materials and Methods for detailed information).  

\begin{figure*}[!t]%
\centering
\includegraphics[width=6in]{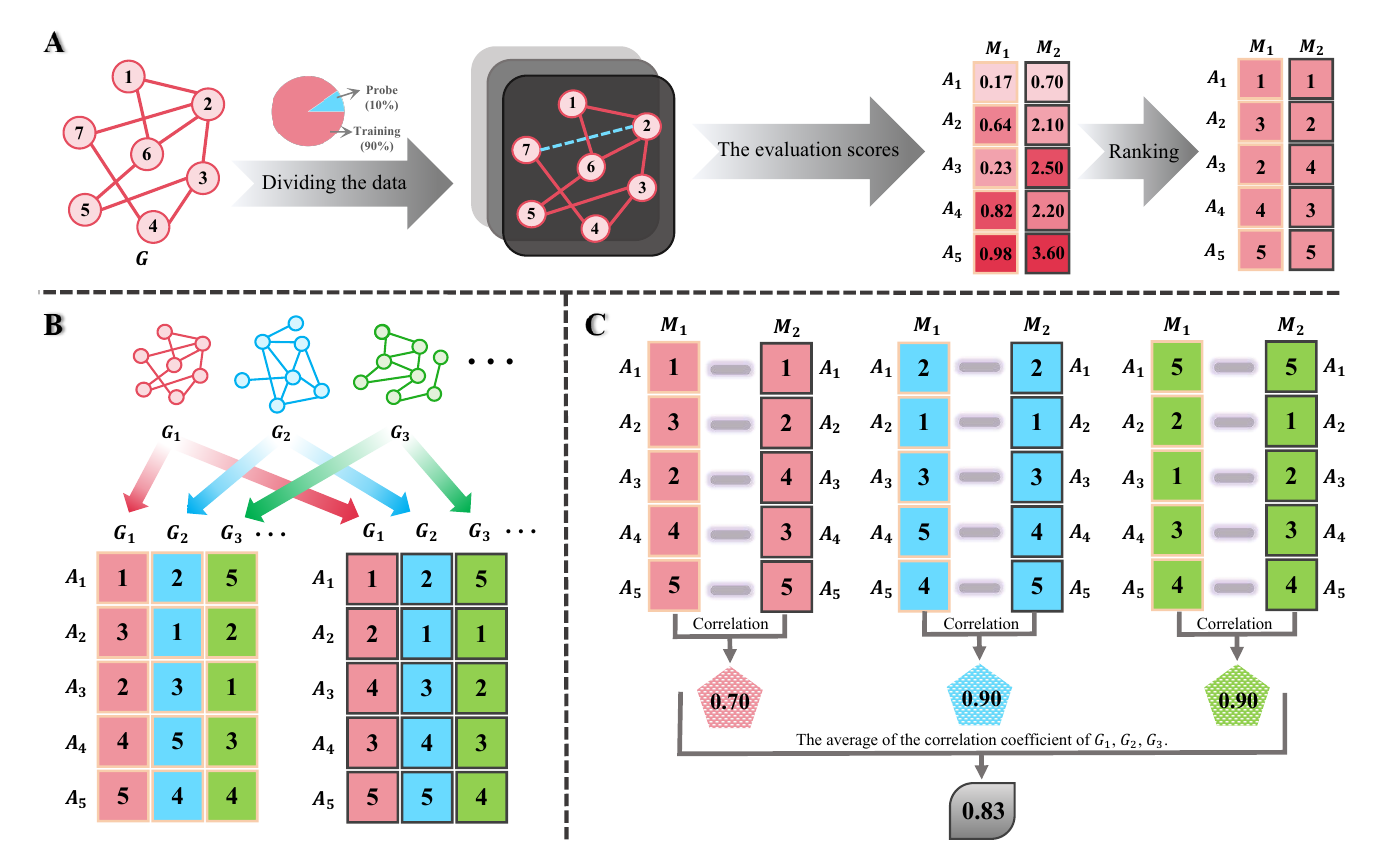}
\caption{Schematic flowchart of the proposed method to measure the correlation between any two evaluation metrics $M_1$ and $M_2$. (A) Initially, the original network is divided into training and probe sets at a ratio, for example 9:1. Next, the evaluation scores of different algorithms $A_{i} (i=1, 2,\dots, P)$ (here we show an example for $P=5$) are calculated by $M_{1}$ and $M_{2}$. The average scores can be obtained by multiple implementations with different random divisions of training and probe sets. Based on the average scores, we can get two rankings of algorithms corresponding to $M_1$ and $M_2$, respectively. (B) We select a large number of real-world networks $G_1, G_2, \cdots, G_Q$, and for each network $G_i$ and each metric $M_j$, we can obtain a ranking of the $P$ algorithms (here we show an example for $Q=3$). (C) We calculate the correlation coefficient of $M_{1}$ and $M_{2}$ by applying some ranking correlation coefficients (e.g., the Spearman correlation coefficient \cite{Spearman1987,wangpei2020} and the Kendall's $\tau$ correlation coefficient \cite{Kendall1938,wangpei2020}) and averaging over the $Q$ selected networks.}\label{fig1}
\end{figure*}

\subsection{Inconsistency among Metrics}
We first calculate the correlations between pairwise metrics by using the above-mentioned framework. The following 12 metrics are under consideration: Precision\cite{buckland1994}, Recall\cite{buckland1994}, Accuracy\cite{swets1988}, Specificity\cite{jones1972}, F1-measure\cite{sasaki2007}, Youden Index\cite{youden1950}, MCC\cite{mattews1975}, AUC\cite{hanely1982}, AUPR\cite{davis2006}, AUC-Precision\cite{carclo2013}, NDCG\cite{jarvelin2002}, and AUC-mROC\cite{carclo2021}. The first seven are threshold-dependent metrics while the last five are threshold-free. The threshold-dependent metrics depend on some threshold parameters, for example, the number of predicted links $k$ (the top-$k$ links with the highest likelihoods are considered as predicted links) or the threshold likelihood $L_c$ (links with likelihoods larger than $L_c$ are considered as predicted links). Detailed definitions of those metrics are presented in the Materials and Methods.

After obtaining the likelihoods of links in $U-E^T$, different kinds of thresholds (e.g., $k$ and $L_{c}$) are essentially equivalent, as the function of any kind of thresholds is to cut all $|U-E^T|$ links into two parts according to their likelihoods. Therefore, we concentrate on the most intuitive and popular threshold $k$. For any fixed $k$, we have rigorously proved that all considered threshold-dependent metrics are equivalent, namely the rankings of algorithms by any two threshold-dependent metrics are exactly the same, provided they share the same $k$. The mathematical proof are shown in the Materials and Methods. As a consequence, we select Precision to represent all threshold-dependent metrics and only discuss Precision later. In addition, since BP is equivalent to Precision at $k=|E^P|$, we will no longer specifically analyze BP. 

\begin{figure*}[!t]%
\centering
\includegraphics[width=5.5in]{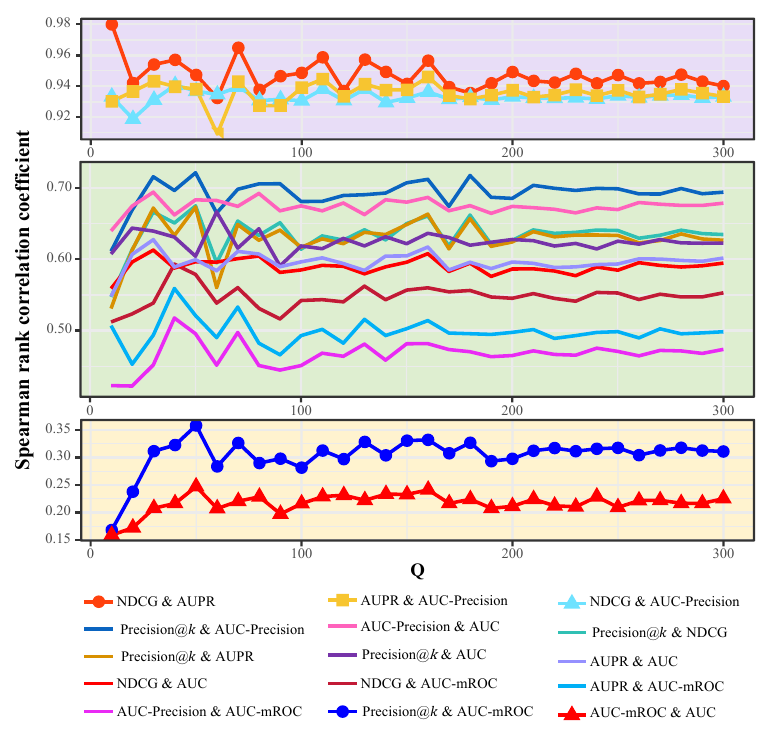}
\caption{The trend of correlations between metrics as the increase of $Q$. For each $Q$, we implement 10 independent runs, where in each run we randomly select $Q$ networks from the collection of 340 real networks. Here the threshold for Precision is set as $k=0.1 \cdot|U-E^{T}|$.}\label{fig2}
\end{figure*}

To obtain the average correlations between metrics, we randomly select $Q$ networks from a collection of 340 real networks in disparate fields, and apply the Spearman correlation coefficient to quantify the correlations. Figure 2 shows the correlations in the range $10\leq Q\leq 300$, where each curve represents a pairwise correlation between two metrics. Obviously, as the increasing of $Q$, the correlation between any two metrics becomes stable. All 15 pairwise relations are divided into three categories: highly correlated (AUPR \& NDCG, AUPR \& AUC-Precision, AUC-Precision \& NDCG), weakly correlated (Precision \& AUC-mROC, AUC \& AUC-mROC), and moderately correlated (others). If two metrics are consistent to each other, their correlation should be close to 1. Unfortunately, as shown in figure 2, except for the three highly correlated metric pairs, the correlations for the other 12 pairs are significantly less than 1, indicating that most pairwise metrics are inconsistent to each other. The results on Kendall's $\tau$ correlation coefficient, and the results for different dividing ratios of training and probe sets are essentially same to the results reported in figure 2 (see SI Appendix).
	
\subsection{Quandary of Threshold-dependent Metrics}
For threshold-dependent metrics, the choices of thresholds are highly relevant. Figure 3 reports how the correlations between Precision and the five threshold-free metrics change for different thresholds $k$. Except for very small $k$ (see the insets of figure 3), all curves exhibit an overall decaying trend as the increasing of $k$, while their decaying patterns are slightly different, namely the Precision-AUC correlation decays slowly and other four curves decay faster. Notice that, every threshold-dependent metric is designed to be meaningful at a relatively small threshold. In contrast, when $k$ approaches its maximum $|U-E^T|$, the score of each metric only depends on the ratio of positive samples to negative samples, irrelevant to the algorithm performance. To summarize, the observed decaying trend results from the fact that Precision (and other threshold-dependent metrics) will become less informative for large $k$. Furthermore, one can infer that if a threshold-free metric puts higher weights to the top-ranked links, the correlation between Precision and this metric will decay faster as the increasing of $k$, because the ranking of links in non-top positions has less effect on this metric. This inference is in line with the observations in figure 3, say the correlation for AUC-mROC decays fastest as AUC-mROC assigns very high weights to top positions by applying logarithmic transformations to both coordinates, and the correlation for AUC decays most slowly since AUC considers the overall advantage of positive samples and is less sensitive to top-ranked links.

As indicated by figure 3, the value of threshold $k$ largely impacts the ranking of algorithms, so how to determine $k$ is still a puzzle needing to be solved. As Precision is originally designed to evaluate the early retrieval performance \cite{herlocker2004,s1_1983}, namely to measure the accuracy of a very few top-ranked predictions, $k$ should be much smaller comparing with the total number of potential links $|U-E^T|$. At the same time, when $k$ is very small, the correlations between Precision and some threshold-free metrics (i.e., AUPR, AUC-Precision and NDCG) are very high, all larger than 0.8. Hence Precision at a very small $k$ provides less additional information to AUPR, AUC-Precision and NDCG. If we choose a large $k$, Precision itself will be less meaningful, though it seems to be more informative at the presence of some threshold-free metrics. Therefore, behind the observations in figure 3 is a quandary in determining the threshold: it should not be small, it should not be large, and it cannot be dug out from the data. In a word, we do not suggest the usage of threshold-dependent metrics if we do not have any clues to determine the threshold. In contrast, if some certain thresholds are meaningful for a specific problem, we can choose one threshold-dependent metric at these thresholds that best fits the practical requirement. For example, if in a e-commercial website, each user will be recommended eight produces (this task can be considered as link prediction in user-product bipartite networks), and the recommender system care most about the click rate, we can choose Precision at $k=8$. 

\begin{figure*}[!t]%
\centering
\includegraphics[width=\textwidth]{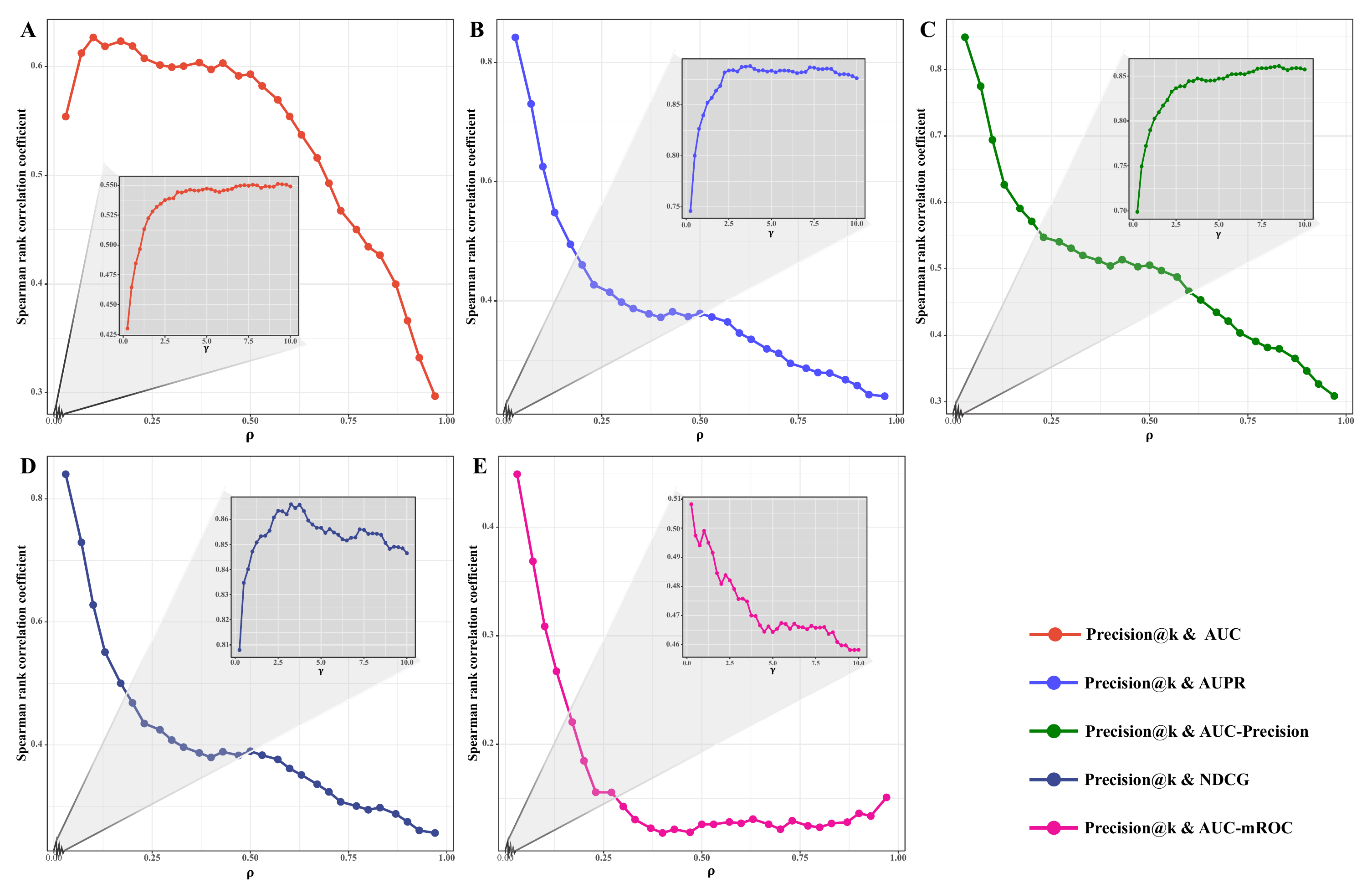}
\caption{The change of correlations between Precision@$k$ and threshold-free metrics for varying $k$. In the main plots, we set $k=\rho|U-E^{T}|$, and in the insets, we set $k=\gamma|E^P|$. The average Spearman rank correlation coefficients correspond to $Q=300$.  (A)-(E) respectively show the cases for AUC, AUPR, AUC-Precision, NDCG, and AUC-mROC.}\label{fig3}
\end{figure*}

\subsection{Correlation Graph Analysis}

\begin{figure*}[!t]%
\centering
\includegraphics[width=5in]{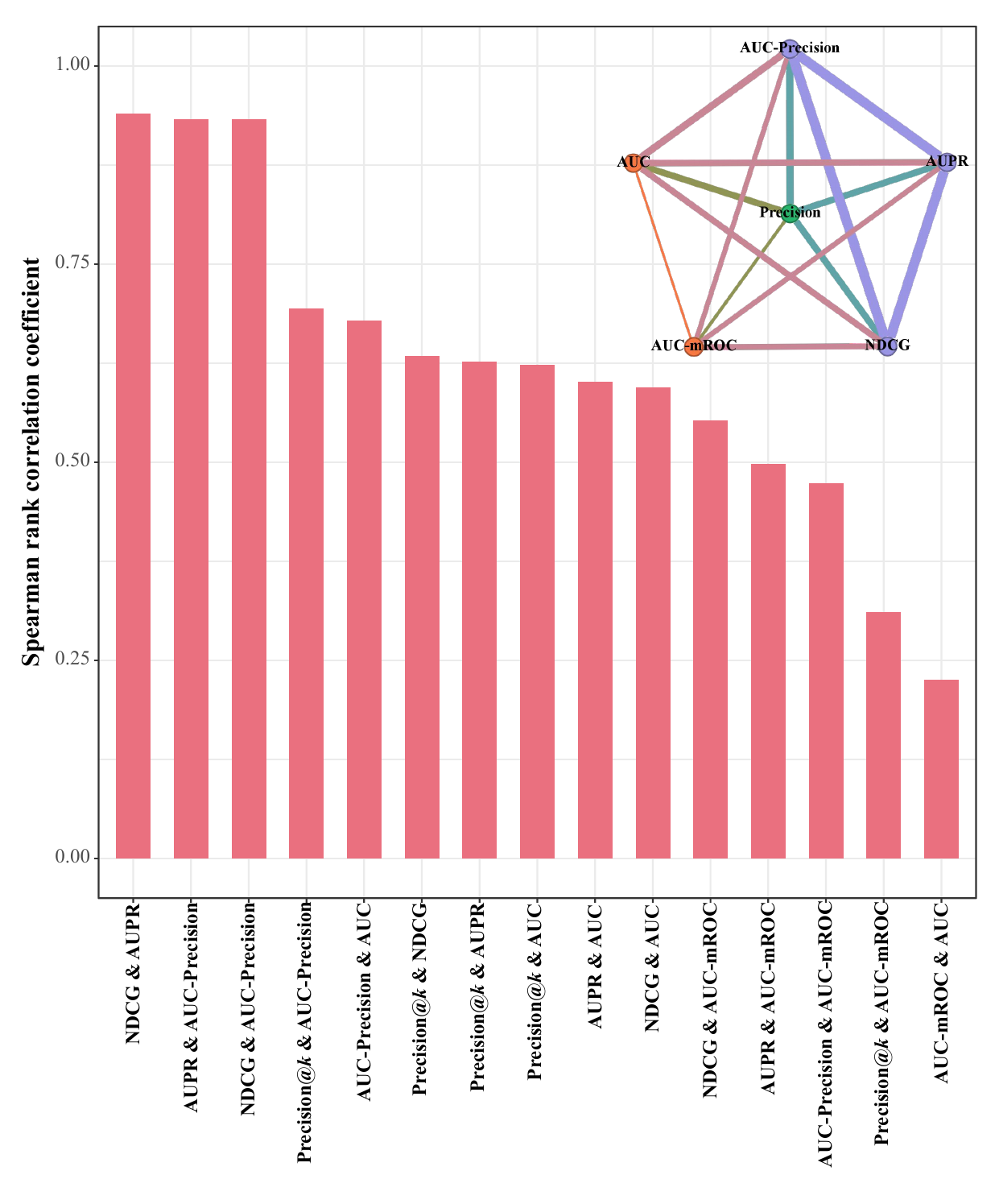}
\caption{The Spearman rank correlation coefficients for all metric pairs, averaged over 10 independent runs and 300 selected networks in each run. For Precision, the threshold is set as $k=0.1\cdot |U-E^{T}|$. The top-right corner shows the corresponding correlation graph, with the thickness of each link representing the strength of correlation. }\label{fig4}
\end{figure*}

To obtain the stable correlations, for each pair of metrics under consideration, we implement 10 independent runs, and for each run we randomly select $Q=300$ networks from the collection of 340 real networks. The average correlations over 300 networks and 10 runs are presented as a histogram in figure 4, ranked in a descending order. The corresponding correlation graph is shown in the top-right corner of figure 4, which is a complete graph (also called clique or fully connected network in the literature) with metrics being nodes and strengths of correlations being link weights. As we have already analyzed the threshold-dependent metrics in the above subsection, here we do not discuss them again but only draw the correlation graph with a specific case $\rho=0.1$. 

The most noticeable structure in the correlation graph is the purple triangle \{AUPR, AUC-Precision, NDCG\}, wherein all pairwise correlations are very high (with an average value 0.936). As a consequence, we suggest only choose one of these three metrics to avoid redundant computation. The correlations between AUC and the above three metrics are of moderate strength, so AUC is still informative even at the presence of some of the three metrics. Therefore, AUC should be considered as one metric for algorithm evaluation. Another conspicuous observation is that AUC-mROC is relatively weakly correlated with other metrics. Indeed, among the 15 pairwise correlations, the least five (i.e., those with the weakest correlations) are all related to AUC-mROC. On the one hand, it is a good point as AUC-mROC is information-rich even at the presence of other metrics, on the other hand, it is a dangerous signal as AUC-mROC will produce a ranking of algorithms probably largely different from all other widely and longly used metrics. This observation is closely related to the individualistic feature of AUC-mROC, namely it pays great attention to a few top-ranked predictions. Therefore, our suggestion is to use AUC-mROC in the scenario where only a limited number of predictions are relevant and the values of those predictions decay fast with their positions. For example, one may search for news, products and friends in some websites, and the search returns are of limited number (the results not appearing in the first page are rarely to be visited) and decaying weights (the click rate decays fast with rank). In such kind of scenarios, AUC-mROC could be relevant.

\section{Discussion}

In this paper, we have implemented extensive experiments involving 340 real networks and 25 algorithms to analyze the consistency among 12 well-known evaluation metrics. As we have proved the essential equivalence of the seven threshold-dependent metrics, our analyses focus on one representative threshold-dependent metric, Precision, and the five threshold-free metrics. Here we emphasize three important observations from the experiments. Firstly, there exists significant and robust inconsistency among evaluation metrics, that is to say, different metrics may provide different rankings of algorithms. Secondly, the ranking of algorithms produced by a threshold-dependent algorithm is sensitive to the threshold $k$, and with the increasing of $k$, the correlation between any threshold-dependent metric and any of the five threshold-free metrics displays an overall decaying trend. The decaying speeds associated with different threshold-free metrics are different: the one with AUC-mROC is the fastest while the one with AUC is the slowest. Thirdly, all pairwise correlations within the set \{AUPR, AUC-Precision, NDCG\} are very high (with an average value 0.936), the correlations between AUC and the above three metrics are of moderate strengths, and the correlations between AUC-mROC and other metrics are weakest.

The above observations are robust to different settings. Firstly, the results are not sensitive to the ratio of training set to probe set. Here, we only report results with $|E^T|:|E^P|=9:1$, while figure 5 (see SI Appendix) shows clearly that such ratio has negligible effect on the metrics' pairwise correlations. Secondly, regarding the ranking correlation coefficients, we consider another famous one, say the Kendall's $\tau$ correlation coefficient. As shown in figure 6 (see SI Appendix), the correlations measures by the Spearman rank correlation coefficient and the Kendall's $\tau$ correlation coefficient exhibit completely the same trend.

For any pair of evaluation metrics, in the calculation of their average correlation over a large number of networks, there are two different methods to aggregate the batch data. The first is to get the ranking of algorithms as well as the corresponding correlation for each network, and them to average those correlations over the $Q$ selected networks. The second is to get the ranking of algorithms for each network at first, and then to obtain a mean rank of each algorithm by averaging its ranks over the $Q$ selected networks, and lastly to calculate the correlation between the two vectors of mean ranks of algorithms produced by the two considered metrics. The first method is more intuitive and thus utilized in this paper, with its detailed procedure presented in figure 1. The schematic flowchart of the second method is shown in figure 7 (see SI Appendix, and only the different part from figure 1 is illustrated). Readers should be aware of that the two methods may result in different relationships between the average correlation and the number of selected networks $Q$. For example, if the average correlation of two metrics $M_1$ and $M_2$ will stabilize at about 0.5 for large $Q$ based on the first method, it does not imply that the average correlation of $M_1$ and $M_2$ will converge in the large limit of $Q$ or will approach to 0.5 if it converges. It is because there exists a rarely observed but mathematically possible situation that two metrics are indeed highly consistent to each other, but their correlation is not very high for a typical network because there is some unknown and unbiased noises that randomly perturb the evaluation scores. Therefore, the correlation for each network is not very high and thus the average correlation obtained by the first method is not very high too. In contrast, the correlation between mean ranks can be very high, because when more and more networks are taken into account, the random effects due to unbiased noises tend to cancel each other out. To be more intuitive, we consider a toy model where two metrics, denoted as $X$ and $Y$, are consistent to each other and both assign an evaluation score $j$ to the $j$th algorithm. However, there exists some random noises that perturb the evaluation scores, say $x_{ij}=j+\sigma_{ij}$ and $y_{ij}=j+\eta_{ij}$ ($i=1, 2,\dots, Q$ and $j=1, 2,\dots, P$) for the $i$th network and $j$th algorithm, where the noises $\sigma_{ij}$ and $\eta_{ij}$ are independently generated from a uniform distribution $U(0, P)$. Figure 8A and figure 8B respectively shows the results obtained by the first and second methods, with $P=100$ is fixed and $Q$ is up to 500. One can observe that, based on the first method, the average correlation will converge to about 0.49 in the large limit of $Q$, while based on the second method, the correlation of mean ranks will approach to 1 for large $Q$, because the random effect vanishes then. As a consequence, we can confidently claim that some evaluation metrics are essentially inconsistent to each other only if both the first and second methods give similar and supportive results. As shown in figure 9 (see SI Appendix), the second method produces qualitatively the same and quantitatively very close results to the first method. Therefore, we can arrive a more believable conclusion that the observed inconsistency among evaluation metrics essentially underlies the definitions of metrics, which does not simply result from some external randomness and thus cannot be eliminated by any statistical skills.     

After extensive experiments and analyses, we eventually arrive to four suggestions about how to select evaluation metrics in link prediction. (i) Despite recent debates, we still recommend AUC as one metric because it has moderate correlations to most metrics and thus can provide additional information to other metrics. (ii) One (and no more than one) of AUPR, AUC-Precision and NDCG should be chosen as a metric. (iii) If we don't have any clues to determine the threshold, it is better not to use threshold-dependent metrics, while if for a specific problem, some thresholds are meaningful, we can choose one (and no more than one) threshold-dependent metric with those thresholds. (iv) To use AUC-mROC in the scenario where only a limited  number of predictions are relevant and the values of those predictions decay fast with their positions.

In general, during the early stages of a discipline's development, exploratory work tends to be more attractive than reflective work, hence the majority of scientists typically allocate their primary efforts to exploring new frontiers. However, once the discipline reaches a certain level of maturity, reflective work becomes essential; otherwise, the defects in the foundation underlying a taller and taller building will lead to greater losses. Link prediction is a young and niche branch of network science. Study in link prediction is very active, with thousands of algorithms being proposed in the past two decades. In comparison, reflective and critical studies are rare \cite{Lichtenwalter2010,Mara2020,carclo2023,zhou2023}. Now is the time for us to reexamine the fundamentals of link prediction research, with a central problem being how to evaluate whether an algorithm is good or bad, or how to compare which of two algorithms performs better. The solution to this problem may be a kind of guideline that we need to follow in the later studies, just as the double-blind principle in medical experiments. Such guideline should clarify at least four issues. (i) \textbf{How to sample the probe set}? It is natural to use random sampling \cite{zhou2011} and temporal sampling \cite{Lichtenwalter2010} to get probe set for missing link prediction and future link prediction, respectively. But there are still some technical details that need to be addressed. For example, how to deal with the situations if the removal of links lead to an unconnected network (this issue becomes more serious for higher-order link prediction \cite{Kumar2020}) or some nodes only appear in the probe set (i.e., all links associated with these nodes are allocated to the probe set). The current approaches are often reasonable yet ad hoc, and thus, we need to assess the extent to which these approaches affect the evaluation of algorithms. For certain specific purposes, there are some other sampling methods. For example, negative sampling method \cite{Kotnis2017} that samples a set of non-existent links with comparable size to the probe set is proposed to manage the cases where the number of missing links is extremely smaller than the number of non-existent links (e.g., for very sparse networks or higher-order networks), and cold sampling method \cite{Zhu2012} that prefers to sample probe links with low-degree ends since in many practical applications to dig out potential interactions between unpopular nodes is more informative and valuable. These less common sampling methods may have subtle but yet unknown relations to algorithm evaluation, meaning that the appropriate methods and metrics for algorithm evaluation can differ under different sampling methods. Very recently, He et al. \cite{he2024} tested 20 different sampling methods and found that different link prediction algorithms exhibit significant differences in accuracy contingent upon the sampling methods. Therefore, the fairness, scope of application and potential impacts of sampling methods requires further analysis and validation. (ii) \textbf{How to determine the model parameters}? An undoubtable principle is any information contained in the probe set cannot be used to train the model parameters. However, in the literature, a commonly-used but incorrect method is to obtain the so-called optimal parameter(s) by comparing the prediction with the probe set. This is largely unfair to parameter-free algorithms, while algorithms that are prone to overfitting will get unjustifiable benefit. Accordingly, on the one hand, in the future studies, researchers should train their model parameters using only the information in $E^T$ (e.g., by further dividing $E^T$ into two parts), on the other hand, maybe more important, we have to reevaluate known algorithms in the above-mentioned fair way. We guess those algorithms that are highly sensitive to parameters and inherently prone to overfitting will exhibit decreased performance, while the relative performance of parameter-free algorithms or those with strong generalization capabilities tends to increase. (iii) \textbf{How to select proper evaluation metrics}? This is a difficult question to answer. While this paper does not provide a complete answer, it raises the value and urgency of the question. The four specific suggestions presented in this paper focus solely on maximizing the informational content of selected metrics without considering the rationality of these metrics themselves. The recommendations may be different if we consider different aspects. For example, if we intend to encourage the early retrieval ability, NDCG and AUC-mROC are good candidates \cite{carclo2021}, while if we emphasize on the discriminating ability, AUC and NDCG are superior \cite{zhou2023,jiao2024}. A fesiable and useful answer may be a combination of suggestions for general tasks and suggestions accounting for some special conditions, such as the data distributions (e.g., the extremely imbalanced learning) and network organization principles (e.g., the higher-order link prediction). (iv) \textbf{How many and how large networks we should use}? In most early studies, only a very few networks (usually of small sizes) are utilized to evaluate the algorithm performance. In comparison, the experiment reported by Ghasemian \textit{et al.} \cite{ghasemian2020} involves 550 real-world networks from diverse fields and of varying sizes. After that, experiments involving a huge number of real-world networks become more popular \cite{Mara2020,Muscoloni2020,Zhou2021}. However, we still lack an analytical or statistical answer as to how many networks and of what size are necessary to obtain a reliable assessment of an algorithm's performance.

This paper only provides a tiny step towards the answer to the third question. However, we believe that the value of this paper is substantial; not only in its provision of four constructive suggestions to help researchers select proper metrics to quickly and accurately evaluate algorithm performance, but also, and perhaps more significantly, it compels us to reevaluate the validity of previously known results. In addition, as link prediction is a kind of binary classification problem, our perspectives and methods could be extended to the selection of evaluation metrics for classification.

\section{Materials and Methods}\label{Materials and Methods}
\subsection{Algorithms of Link Prediction}\label{Algorithms}
In this work, we consider 25 algorithms. Some are well-known and some are very recently proposed. Table 1 lists those algorithms, together with the corresponding references, where readers can find more details.
	\begin{table*}[h]
	\centering
	\caption{List of link prediction algorithms considered in this paper, with abbreviations showing in the brackets.}\label{tab1}
	\begin{tabular}{cc}
		\hline
		\textbf{Algorithms} & \textbf{References} \\
		\hline
		Common Neighbor Index (CN)    & \cite{cn_2001} \\
		Resource Allocation Index (RA)    & \cite{zhou2009}\\
		Local Path Index (LP)    & \cite{zhou2009}\\
		Adamic-Adar Index (AA)    & \cite{aa_2003}  \\
		Preferential Attachment Index (PA)    & \cite{pa_2002}  \\
		Jaccard Index    & \cite{jc_1901}  \\
		Average Commute Time (ACT)    & \cite{act_2010} \\
		Sim Index    & \cite{sim_2020}  \\
		Length Three (L3)    & \cite{kovacs2019}  \\
		Adjacency Three (A3)   & \cite{kovacs2019}  \\
		Katz Index & \cite{katz_1953}\\
		Liner Optimization (LO)    &  \cite{lo_2019}\\
		Salton Index   & \cite{s1_1983}  \\
		Sørenson Index   & \cite{so_1948} \\
		Hub Promoted Index (HPI)   &   \cite{hpi_2002}\\
		Hub Depressed Index (HDI)  &  \cite{zhou2011} \\
		Local Random Walk (LRW)  &  \cite{liu2010} \\
		Superposed Random Walk (SRW)   &  \cite{liu2010} \\
		Leicht-Holme-Newman-1 Index (LHN-1)    &  \cite{lhn_2006}\\
		Matrix Forest Index (MFI) & \cite{mfi_1997}  \\  
		Local Naive Bayes based Adamic-Adar Index (LNBAA)    &  \cite{bayes_2011} \\   
		Local Naive Bayes based Resource Allocation Index (LNBRA)   &  \cite{bayes_2011} \\
		Salton Cosine Similarity (S1) &  \cite{s1_1983} \\
		Controlling the Leading Eigenvector (CLE) & \cite{cle_2021}  \\
		Common neighbor and Centrality based Parameterized Algorithm (CCPA) & \cite{ccpa_2020}  \\
		\hline
	\end{tabular}
	\end{table*}

\subsection{Evaluation Metrics}\label{metrics}
This subsection will introduce the 12 considered metrics, say Precision, Recall, Accuracy, Specificity, F1-measure, Youden, MCC, AUC, AUPR, AUC-Precision, NDCG, and AUC-mROC. The first seven are threshold-dependent metrics and the last five are threshold-free metrics. 

Without loss of generality, we assume that each algorithm will assign a score $s_{ij}$ to characterize the existence likelihood of any potential link $(i,j) \in U-E^{T}$, and all links in $U-E^{T}$ are ranked in a descending order of their scores. The threshold $k$ cuts the set of potential links into two parts: the top-$k$ ranked links are predicted missing links, while the others are predicted non-existent links. As link prediction is a binary classification problem, we can use the confusion matrix to formulate threshold-dependent metrics. In the confusion matrix, all samples are classified into four categories based on whether they are positive samples or negative samples, and whether they are correctly predicted. These four categories are: true positive (TP), where a positive sample is correctly predicted as positive; false positive (FP), where a negative sample is incorrectly predicted as positive; true negative (TN), where a negative sample is correctly predicted as negative; and false negative (FN), where a positive sample is incorrectly predicted as negative.

Next we can describe the threshold-dependent metrics using the language of confusion matrix. \textbf{Precision} is proportion of true positives to all predicted positives \cite{buckland1994}. \textbf{Recall} measures the ratio of true positives to the total number of positives \cite{buckland1994}. \textbf{Accuracy} quantifies the proportion of correctly classified instances out of the total instances \cite{swets1988}. \textbf{Specificity} measures the ratio of true negatives to the total number of negatives \cite{jones1972}. \textbf{F1-measure} is the harmonic mean of Precision and Recall \cite{sasaki2007}. \textbf{Youden Index} is defined as the sum of Recall and Specificity minus 1, which captures the overall performance of a diagnostic test \cite{youden1950}. \textbf{MCC} takes into account the roles of all elements in the confusion matrix, which is  particularly useful when dealing with imbalanced learning problem \cite{mattews1975}. Accordingly, the mathematically formulas for these threshold-dependent metrics are as follows.

	\begin{equation}
		Precision=\frac{TP}{TP+FP}=\frac{TP}{k}, 
		\label{Precision}
	\end{equation}
	
	\begin{equation}
		Recall=\frac{TP}{TP+FN}=\frac{TP}{|E^P|},
		\label{Recall}
	\end{equation}
	
	\begin{equation}
		Accuracy=\frac{TP+TN}{TP+FP+FN+TN}=\frac{TP+TN}{|U-E^T|},  
		\label{Accuracy}
	\end{equation}
	
	\begin{equation}
		Specificity=\frac{TN}{TN+FP}=\frac{TN}{|U-E|}, 
		\label{Specificity}
	\end{equation}
	
        \begin{equation}
		F1=2\left(\frac{1}{Precision}+\frac{1}{Recall}\right)^{-1}=\frac{2\cdot Precision\cdot Recall}{Precision + Recall},\label{F1}
	\end{equation}
	
	\begin{equation}
		Youden=Recall+Specificity-1, 
		\label{Youden}
	\end{equation}
	
	\begin{equation}
		MCC=\frac{TP\cdot TN - FP \cdot FN }{\sqrt{(TP+FP)(TP+FN)(TN+FP)(TN+FN)}}.
		\label{MCC}
	\end{equation}

\textbf{AUC} measures the ability of the model to discriminate between positive and negative classes across all possible threshold values. It delegates the are under the Receiver Operating Characteristic (ROC) curve \cite{hanely1982}. The range of AUC is $[0,1]$, where a higher value indicates better performance. As AUC is equavalent to the probability that a randomly selected positive sample (i.e., missing link) is scored higher than a randomly selected negative sample (i.e., non-existent link), we can obtain the approximation of AUC through directly comparing positive and negative samples. If we randomly compare $n$ positive-negative pairs, and  there are $n_{1}$ times the missing link having higher score and $n_{2}$ times the missing link and nonexistent link having the same score, then AUC is approximated as
\begin{equation}
AUC =\frac{n_{1} + 0.5 n_{2}}{n}.
\label{AUC}
\end{equation}
AUC will approach 0.5 if the algorithm produce a random classification, and thus to what extent AUC exceeds 0.5 indicates how much better the algorithm performs than pure chance. \textbf{AUPR} measures the area under the precision-recall curve, which plots Precision (on the Y-axis) against Recall (on the X-axis) for different threshold values \cite{davis2006}. A higher AUPR value indicates better performance. If the positions of the $|E^{P}|$ missing links are $r_{1}<r_{2}<\dots<r_{|E^{P}|}$ in the $|U-E^{T}|$ ranked links, then AUPR can be calculated as
\begin{equation}
AUPR=\frac{1}{2|E^{P}|} {\left ( \sum_{i=1}^{|E^{P}|} \frac{i}{r_{i}} +\sum_{i=1}^{|E^{P}|} \frac{i}{r_{i+1}-1}
\right )} , 
\label{AUPR}
\end{equation}
where $r_{|E^{P}|+1}$ is defined as $|U-E^{T}|+1$. Analogously, \textbf{AUC-Precision} is the area under the threshold-precision curve, which plots Precision (on the Y-axis) against the threshold $k$ (on the X-axis) for different threshold values \cite{carclo2013}. \textbf{NDCG} assigns larger weights to higher positions, normalized by the ideal discounted cumulative gain, as \cite{jarvelin2002}
\begin{equation}
NDCG=\sum_{i=1}^{\left | E^P  \right | } \frac{1}{\log_{2}{(1+r_i)}} 
\bigg/ \sum_{l=1}^{\left | E^P  \right | } \frac{1}{\log_{2}{(1+l)}},
\label{NDCG}
\end{equation}
where the contribution of a missing link at position $r$ is $\frac{1}{\log_{2}(1+r)}$. \textbf{AUC-mROC} is a variant of AUC by transforming both axes of the ROC curve using logarithmic scale \cite{carclo2021}. It applies another transformation to ensure the a random classification also lies in the diagonal line. The finalized horizontal and vertical coordinates are defined as 
\begin{equation}
mFPR= \log_{J}{(1+FP)}
\end{equation}
and
\begin{equation}
mTPR= mFPR+[\log_{Z}{(1+TP) - H}] \cdot (1-H)^{-1} \cdot (1-mFPR),
\end{equation}
where $J=1+|U-E|$, $Z=1+|E^{P}|$, and $H=\log_{Z}{(1+FP \cdot \frac{Z-1}{J-1})}$. The AUC-mROC is the area under the above transformed curve.

\subsection{Equivalence of Threshold-dependent Metrics} \label{proof}
\textbf{Theorem.} 
\textit{Given the threshold $k$, metrics in the set $\Omega$=\{Precision, Recall, F1, Specificity, Youden, Accuracy, MCC\} are equilevent to each other, namely any two metrics in $\Omega$ will give exactly the same rankings of algorithms.}

\begin{proof}
Consider two link prediction algorithms $A_1$ and $A_2$, and any two metrics $M_{i}, M_{j} \in \Omega$, denote $M_i(A_1)$ the evaluation score $A_1$ received from $M_i$, then the theorem can be unfolded to the following three propositions: (1) if $M_{i}(A_{1}) < M_{i}(A_{2})$, then $M_{j}(A_{1}) < M_{j}(A_{2})$; (2) if $M_{i}(A_{1}) > M_{i}(A_{2})$, then $M_{j}(A_{1}) > M_{j}(A_{2})$, (3) if $M_{i}(A_{1}) = M_{i}(A_{2})$, then $M_{j}(A_{1}) = M_{j}(A_{2})$. In subsequent proof, it is assumed that $k$ is given as a constant.

Obviously, from proposition (1), we can deduce propositions (2) and (3): (1)$\Rightarrow$(2) can be obtained by exchanging $A_1$ and $A_2$, (1)$\Rightarrow$(3) can be proved by contradiction. Therefore, to prove the equivalence between two metrics $M_i$ and $M_j$, we only need to show that for any two algorithms $A_1$ and $A_2$, if $M_{i}(A_{1}) < M_{i}(A_{2})$, then $M_{j}(A_{1}) < M_{j}(A_{2})$. For convenience, we use Precision as the central metric, and then prove the following six inequalities provided the condition $Precision(A_1)<Precision(A_2)$: (i) $Recall(A_{1}) < Recall(A_{2})$; (ii) $F1(A_{1}) < F1(A_{2})$; (iii) $Specificity(A_{1}) < Specificity(A_{2})$; (iv) $Youden(A_{1}) < Youden(A_{2})$; (v) $Accuracy(A_{1}) < Accuracy(A_{2})$; (vi) $MCC(A_{1}) < MCC(A_{2})$.

According to the definition in Eq. (1), as $k$ is fixed, the condition $Precision(A_{1}) < Precision(A_{2})$ is equivalent to $TP(A_{1}) < TP(A_{2})$, so that a smart way to prove the above inequalities is expressing elements in the confusion matrix by $TP$ and other constants. Using the following evident relationships  
\begin{equation}
    \begin{aligned}
    & FP+TP=k, \\
    & FN+TP=|E^{P}|, \\
    & TN+FP=|U-E|, \\ 
    & TN+FN=|U-E^{T}|-k, \\ 
    \end{aligned}
    \label{Confusion}
\end{equation}
we have 
\begin{equation}
    \begin{aligned}
    & FP=k-TP, \\
    & FN=|E^{P}|-TP, \\
    & TN=|U-E|-k+TP. \\  
    \end{aligned}
\end{equation}

Next, we prove the six inequalities one by one. Inequality (i) is evident as 
\begin{equation}
Recall(A_1)=\frac{TP(A_1)}{|E^P|}<\frac{TP(A_2)}{|E^P|}=Recall(A_2).
\end{equation}
If $Precision(A_{1}) < Precision(A_{2})$ and $Recall(A_{1}) < Recall(A_{2})$, it is very clear that the harmonic mean of $Precision(A_1)$ and $Recall(A_1)$ is also smaller than the harmonic mean of $Precision(A_2)$ and $Recall(A_2)$, say the inequality (ii) holds. Substituting Eq. (14) to Eq. (4), we have 
\begin{equation}
Specificity(A_1)=\frac{|U-E|-k+TP(A_1)}{|U-E|}<\frac{|U-E|-k+TP(A_2)}{|U-E|}=Specificity(A_2),
\end{equation}
namely the inequality (iii) holds.
If $Precision(A_{1}) < Precision(A_{2})$, we can deduce that $Recall(A_{1}) < Recall(A_{2})$ and $Specificity(A_{1}) < Specificity(A_{2})$ by inequality (i) and inequality (iii) respectively, hence $Youden(A_1)<Youden(A_2)$ according to the definition Eq. (6), namely the inequality (iv) holds. Combining Eq. (3) and Eq. (14), we have
\begin{equation}
Accuracy=\frac{|U-E|-k+2TP}{|U-E^T|},
\end{equation}
so that
\begin{equation}
Accuracy(A_1)=\frac{|U-E|-k+2TP(A_1)}{|U-E^T|}<\frac{|U-E|-k+2TP(A_2)}{|U-E^T|}=Accuracy(A_2),
\end{equation}
namely the inequality (v) holds. According to Eq. (14), the numerator of MCC is 
\begin{equation}
TP\cdot TN-FP\cdot FN=TP(|U-E|-k+TP)-(k-TP)(|E^P|-TP)=(|U-E^T|TP-k|E^P|),
\end{equation}
and the denominator of MCC can be expressed by Eq. (13), therefore
\begin{equation}
MCC=\frac{|U-E^T|TP-k|E^P|}{k|E^{P}||U-E|(|U-E^{T}|-k)},
\end{equation}
and thus 
\begin{equation}
MCC(A_1)=\frac{|U-E^T|TP(A_1)-k|E^P|}{k|E^{P}||U-E|(|U-E^{T}|-k)}<\frac{|U-E^T|TP(A_2)-k|E^P|}{k|E^{P}||U-E|(|U-E^{T}|-k)}=MCC(A_2),
\end{equation}
namely the inequality (vi) holds.
\end{proof}

\subsection{Ranking Correlation Coefficients}\label{Similarity}
This paper applies two classical coefficients to measure the correlation between two rankings, say Spearman rank correlation coefficient \cite{Spearman1987, wangpei2020} and Kendall's $\tau$ correlation coefficient \cite{Kendall1938, wangpei2020}. Denoting $R_{iu}$ the rank of the $u$th algorithm by the $i$th metric, the Spearman rank correlation coefficient between two rankings produced by metrics $i$ and $j$ is
\begin{equation}
    r_{ij}=\frac{\sum_{u=1}^{P}(R_{iu}-\bar{R}_i)(R_{ju}-\bar{R}_j)}{\sqrt{\sum_{u=1}^{P}(R_{iu}-\bar{R}_i)^{2}}\cdot \sqrt{\sum_{u=1}^{P}(R_{ju}-\bar{R}_j)^{2}}},
\end{equation}
where $P=25$ is the number of algorithms under consideration and $\bar{R}_{i}$ is the average rank. Clearly, the Spearman rank correlation coefficient lies in the range $-1\leq r_{ij}\leq 1$. The Kendall's $\tau$ measurs the strength of association of the cross tabulations. Considering two algorithms $u$ and $v$ ($1\leq u,v\leq P$) and two metrics $i$ and $j$, if $R_{iu}>R_{iv}$ and $R_{ju}>R_{jv}$, or $R_{iu}<R_{iv}$ and $R_{ju}<R_{jv}$, we say the pair $(u,v)$ is concordant, if $R_{iu}>R_{iv}$ but $R_{ju}<R_{jv}$, or $R_{iu}<R_{iv}$ but $R_{ju}>R_{jv}$, we say the pair $(u,v)$ is discordant, and if $R_{iu}=R_{iv}$ or $R_{ju}=R_{jv}$, we say the pair $(u,v)$ is tied. Counting all $P(P-1)/2$ pairs, the Kendall's $\tau$ reads
\begin{equation}
    \tau_{ij}=\frac{2(N_{C}-N_{D})}{P(P-1)},
\end{equation}
where $N_{C}$ is the number of concordant pairs, and $N_{D}$ is the number of discordant pairs.

\subsection{Data and Codes}
The 340 real-world networks mainly come from the two public datasets (\textcolor{blue}{http://konect.cc/networks/} and \textcolor{blue}{https://networkrepository.com/networks.php}). The details of the data and code are deposited in GitHub: \textcolor{blue}{https://github.com/98YiLin/IEMLP.git}.

\bibliographystyle{plain}

\newpage
\section{Supplemental Information} \label{SI}
\subsection{Sensitivity Analysis}\label{Setting}
\begin{figure*}[!htbp]%
    \centering
    \includegraphics[width=3.0in]{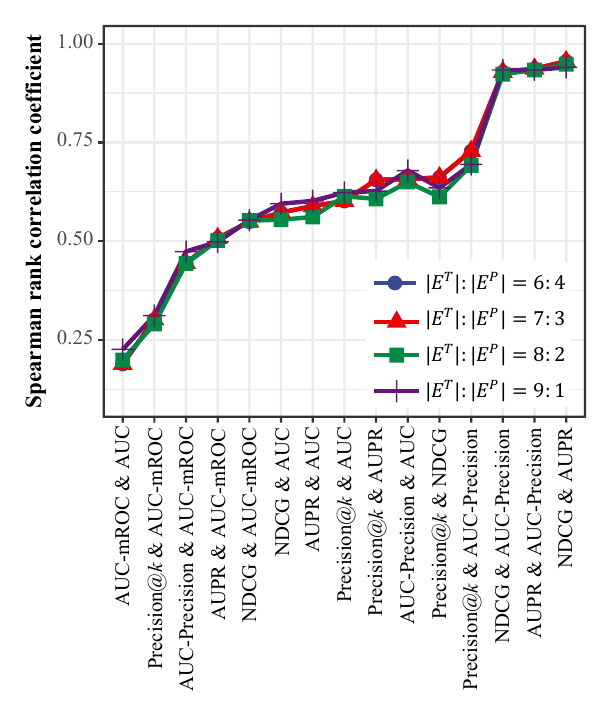}
    \caption {The average pairwise correlations over 300 randomly selected real networks for different splitting ratios of $|E^T|$ to $|E^P|$. The blue, red, green, and purple lines represent the results for $|E^T|:|E^P|=6:4$, $|E^T|:|E^P|=7:3$, $|E^T|:|E^P|=8:2$, and $|E^T|:|E^P|=9:1$, respectively.}\label{SI_Ratio}
    \end{figure*}

\begin{figure*}[!htbp]%
    \centering
    \includegraphics[width=3.0in]{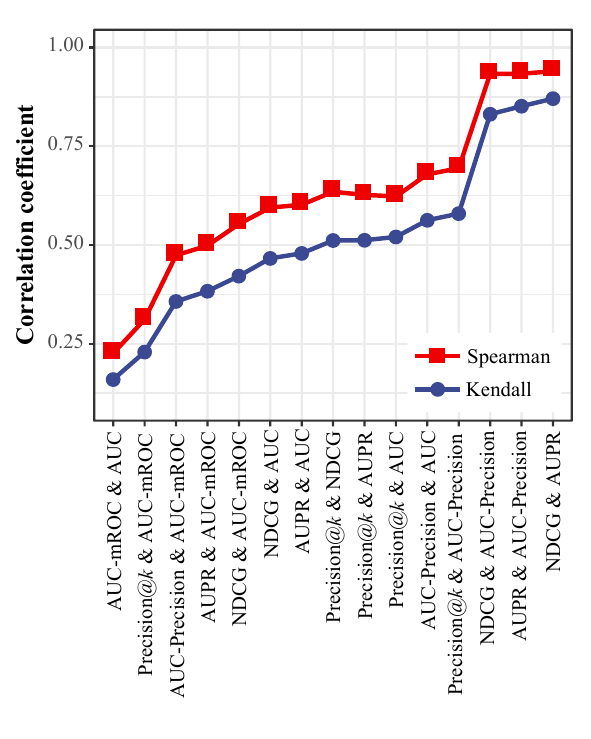}
    \caption {The average pairwise correlations over 300 randomly selected real networks, obtained by different correlation coefficients. The red and blue lines represent the results obtained using the Spearman rank correlation coefficient and the Kendall's $\tau$, respectively.}\label{SI_KCC}
    \end{figure*}

We first test the impacts of the ratio of the training set to the probe set. In addition to the commonly used ratio \cite{zhou2011}, say $|E^T|:|E^P|=9:1$, we consider other ratios like 8:2, 7:3 and 6:4, which are also usually used in binary classification. As shown in figure \ref{SI_Ratio}, the change in ratio does not affect the correlations between metrics, suggesting the robustness of observations in the main text. We next check whether our results are sensitive to the choice of correlation measures by comparing the Spearman rank correlation coefficient with another well-known coefficient, say the Kendall's $\tau$. As shown in figure \ref{SI_KCC}, they show completely the same trend, indicating that our results are robust to the correlation coefficients.

\subsection{The Alternative Method } \label{AnotherMethod}
\begin{figure*}[!t]%
    \centering
    \includegraphics[width=4in]{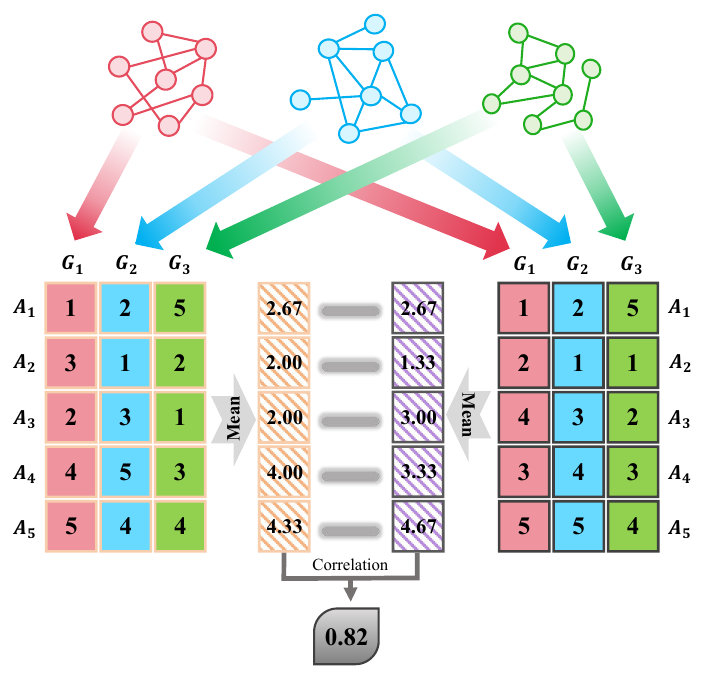}
    \caption {Schematic flowchart of an alternative averaging method to measure the correlation between any two evaluation metrics $M_1$ and $M_2$ in for five algorithms ($P=5$). After obtaining the rankings of algorithms for the $Q$ selected networks (here we show an example for $Q=3$), we first calculate the mean ranks and then measure the correlation between two vectors of mean ranks.}\label{fig1SI}
    \end{figure*}

Different from the method in figure \ref{fig1}, there is an alternative way to calculate the correlation between metrics based on a large number of real networks. The key point of this method is to first average the ranks of different algorithms over selected networks, and then calculate the Spearman rank correlation coefficient of the mean ranks. The different part of this method from the method applied in the main text is shown in figure \ref{fig1SI}. Figure 8 shows the results of the toy model introduced in the main text by using the first and second methods. For the first method (see figure 8A), when $Q$ is large enough, the correlation coefficient between $X$ and $Y$ stabilizes at about 0.49. However, for the second method (see figure 8B), the correlation coefficient rapidly increases to 1 as $Q$ increases. Figure 9 reports the results by using the second method for real networks, where the other settings are completely the same to those of figure 4. 

    \begin{figure*}[!htbp]%
    \centering
    \includegraphics[width=5in]{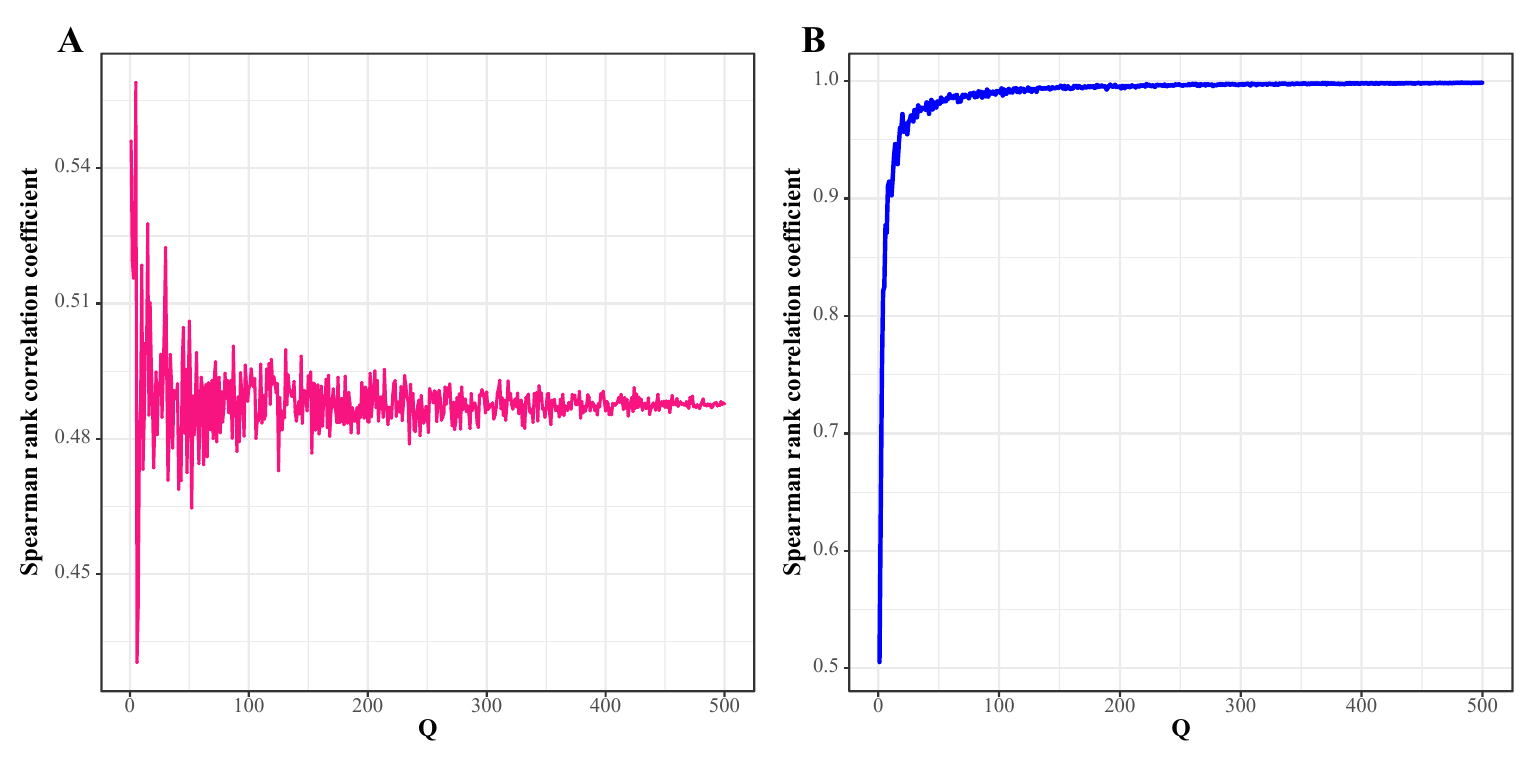}
    \caption {The Spearman rank correlation coefficients between $X$ and $Y$ as the increasing of $Q$ for the toy model using (A) the first method and (B) the second method.}\label{SI_method}
    \end{figure*}
    
    \begin{figure*}[!t]%
    \centering
    \includegraphics[width=3.6in]{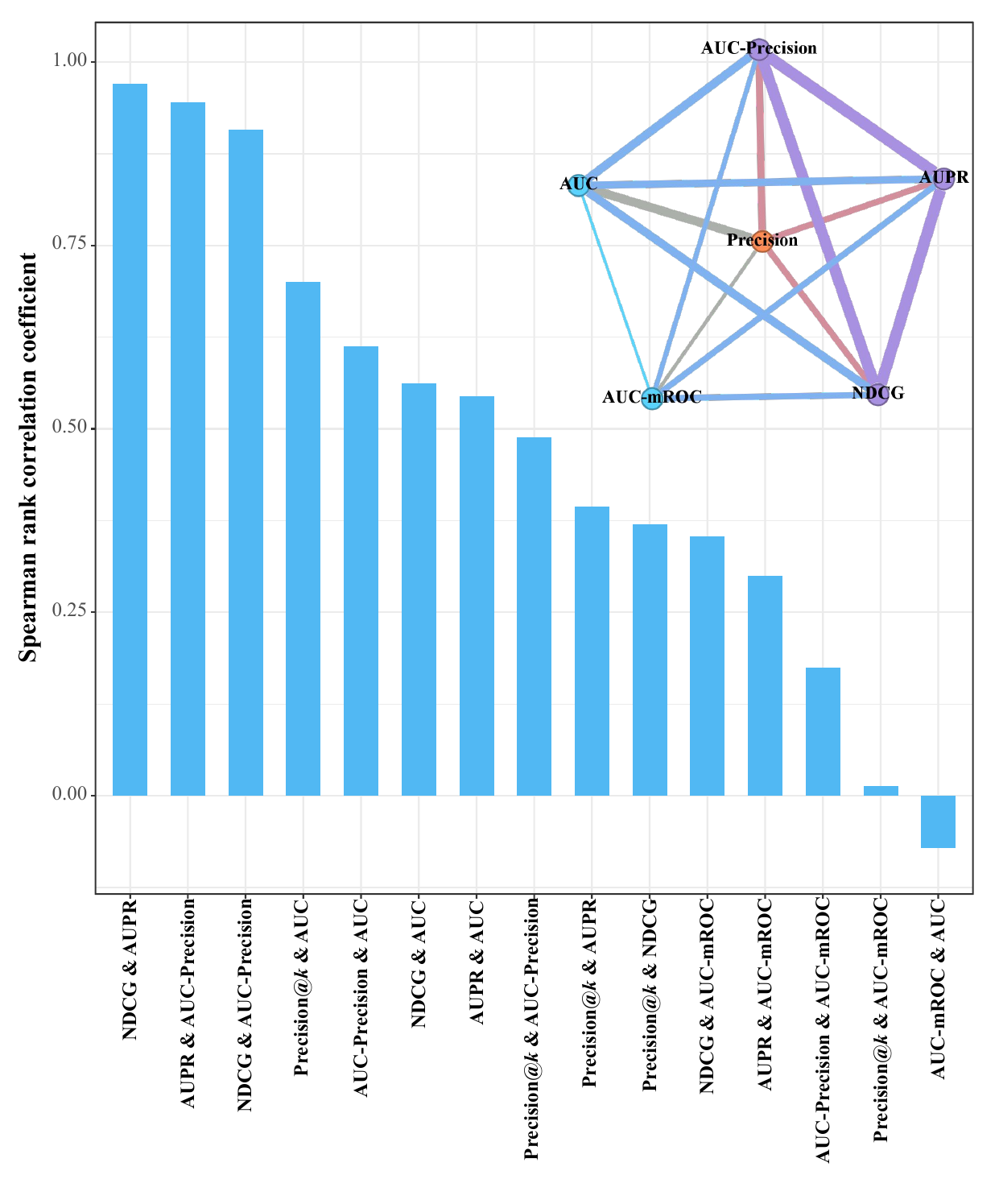}
    \caption {The Spearman rank correlation coefficients for all metric pairs, obtained by the method presented in figure \ref{fig1SI}, which are averaged over 10 independent runs and 300 selected networks in each run. For Precision, the threshold is set as $k=0.1\cdot |U-E^{T}|$. The top-right corner shows the corresponding correlation graph, with the thickness of each link representing the strength of correlation.}\label{fig4SI}
    \end{figure*}	 

\end{document}